# The Impact of Artificial Space Objects on Optical Astronomy as Determined from 32 Million Photometric Observations

Anthony Mallama [1], Sergey Karpov [2] and Richard E. Cole [1]

[1] IAU Centre for the Protection of the Dark and Quiet Sky,
IAU-UAI Headquarters, 98-bis Blvd Arago, 75014,
Paris, France

[2] Institute of Physics, Czech Academy of Sciences,
CZ-182 21 Prague 8, Czech Republic

Correspondence: anthony.mallama@gmail.com

Photometric observations of satellites, rockets and space debris brighter than magnitudes 6.0, 7.0 and 8.0 are characterized. Those magnitudes pertain to different levels of interference with astronomy as defined by the International Astronomical Union. The densities of objects per square degree of sky and brighter than the magnitudes listed above are 0.38 e-3, 0.76 e-3 and 1.10 e-3, respectively. Probabilities of objects appearing on astronomical images depend on the field of view and the exposure duration. For objects brighter than magnitude 7.0 the probability that one will appear on a 10 second exposure of a 10 degree square field of view is 15%. The probability that an object is visible to the unaided eye under dark sky conditions is 59%. These percentages represent all nighttime hours. We discuss how they change with the Sun's distance below the horizon. Examination of data for objects launched since the start of the spacecraft constellation era shows that satellites account for most of the bright observations. If the plans of satellite operators to launch a million spacecraft are realized, nearly every astronomical image of a few seconds duration and several degrees in size may be contaminated during several hours of each night. Finally, people viewing the night sky may perceive satellites more strongly than stars.

## 1. Introduction

Bright space objects interfere with astronomical observations (Barentine et al. 2023) and spoil the aesthetic beauty of the night sky (Mallama and Young 2021). This problem has worsened with the launch of large satellite constellations.

Brightness statistics for such objects have been reported by Mallama and Cole (2025), and Mallama et al (2026a and 2026b). That research used magnitudes recorded by the Mini-MegaTORTORA (MMT9) robotic observatory (Karpov et al. 2016; Beskin et al. 2017) supplemented with visual observations. Our results agree closely with values for three versions of Starlink satellites derived by Zhi et al (2024). Other recent studies including that by Longa-Peña et al (2026) report similar brightness characteristics.

This paper reports on satellites, rockets and debris objects recorded during the year 2025. Our analysis indicates their current and future impact on optical astronomy.

Section 2 discusses three levels of brightness that interfere with astronomy. Section 3 describes the MMT9 system and its database of observations. Section 4 discusses three types of objects: satellites, rockets and debris. Section 5 presents the magnitude distributions of these objects. Section 6 reports on the densities of objects per square degree of sky. Section 7 lists probabilities that the objects will appear on astronomical images. Section 8 gives probabilities that they will be seen by the unaided eye. Section 9 compares the impact of the three types of objects launched before and after the beginning of the satellite constellation era. Section 10 discusses how the visibility of space objects changes during the hours of darkness, the future impact of artificial space objects and the findings of related studies. Section 11 summarizes our results.

## 2. Brightness limits for astronomy

The International Astronomical Union (IAU, 2024) defined upper limits to brightness that minimize interference with observational research. For satellite altitudes up to 550 km that limit is Johnson visual magnitude, $M_V$, 7.0. Equation 1 applies above that height,

$$M_V = 7.0 + 2.5 * \log10 ( altitude / 550 )$$

Equation 1

where *altitude* refers to the satellite's height above sea level in km. The equation evaluates to $M_V$ = 8.0 for an altitude of 1,380 km which exceeds the height for most objects orbiting the Earth.

The IAU statement also recommends that spacecraft should not be visible to the unaided eye. Objects of visual magnitude 6.0 can be seen at locations where the sky is minimally affected by light pollution. The magnitude 6, 7 and 8 limits are used to assess the impact of satellites, rockets and debris on astronomy.

## 3. Observations

Photometric measurements are recorded by the Mini-MegaTORTORA (MMT9) robotic observatory in Russia at 43.65N and 41.43E (Karpov et al. 2016; Beskin et al. 2017). The observatory has been operating since 2014 and its magnitudes are within 0.1 of the Johnson V-band as discussed by Mallama (2021).

MMT9 acquires satellite data in tracks across the sky at 10 Hz frequency. We extracted from the online database of MMT9 satellite observations (Karpov et al. 2016) a set of 32 million magnitudes for 11,241 objects in the NORAD (North American Aerospace Defense Command) catalog that were recorded during 2025. A limitation of this study is that all observations were recorded at one location. This single site does not sample all orbital populations equally.

The publicly available portion of the MMT9 database does not include Russian objects. So, we estimated the numbers of their magnitudes as described in the next section.

**4. Object types and time periods**

Rocket bodies are identified in the NORAD catalog by the letters 'RB' in their names, while debris object names contain 'DEB'. All other bodies are satellites. The most numerous objects in the NORAD catalog are satellites, followed by debris and rockets.

We derived a correction factor to account for the missing Russian observations in the MMT9 database as follows. Every twentieth object in the NORAD catalog was sampled individually, noting the country of origin and the type of object. We also recorded whether the object was launched before or after 2020.0 which we take to be the beginning of the large satellite constellation era. Results for the 1,460 samples are listed in Table 1 and plotted in Figure 1.

The correction factor to convert from magnitude counts for non-Russian objects to all objects is given by Equation 2,

$$C = 1 + ( R / N ) \qquad \text{Equation 2}$$

where $C$ is the correction, $R$ is the number of Russian objects and $N$ is the number of non-Russian objects. The factors listed in Table 1 are applied to all the observations discussed below.

Table 1. NORAD Catalog Analysis

| Year | Russ | Not | Corr | +/- |
|---|---|---|---|---|
| **Satellites** | | | | |
| < 2020.0 | 69 | 141 | 1.49 | 0.10 |
| > 2020.0 | 25 | 603 | 1.04 | 0.01 |
| **Rockets** | | | | |
| < 2020.0 | 41 | 51 | 1.80 | 0.24 |
| > 2020.0 | 2 | 10 | 1.20 | 0.18 |
| **Debris** | | | | |
| < 2020.0 | 158 | 296 | 1.53 | 0.07 |
| > 2020.0 | 1 | 61 | 1.02 | 0.02 |

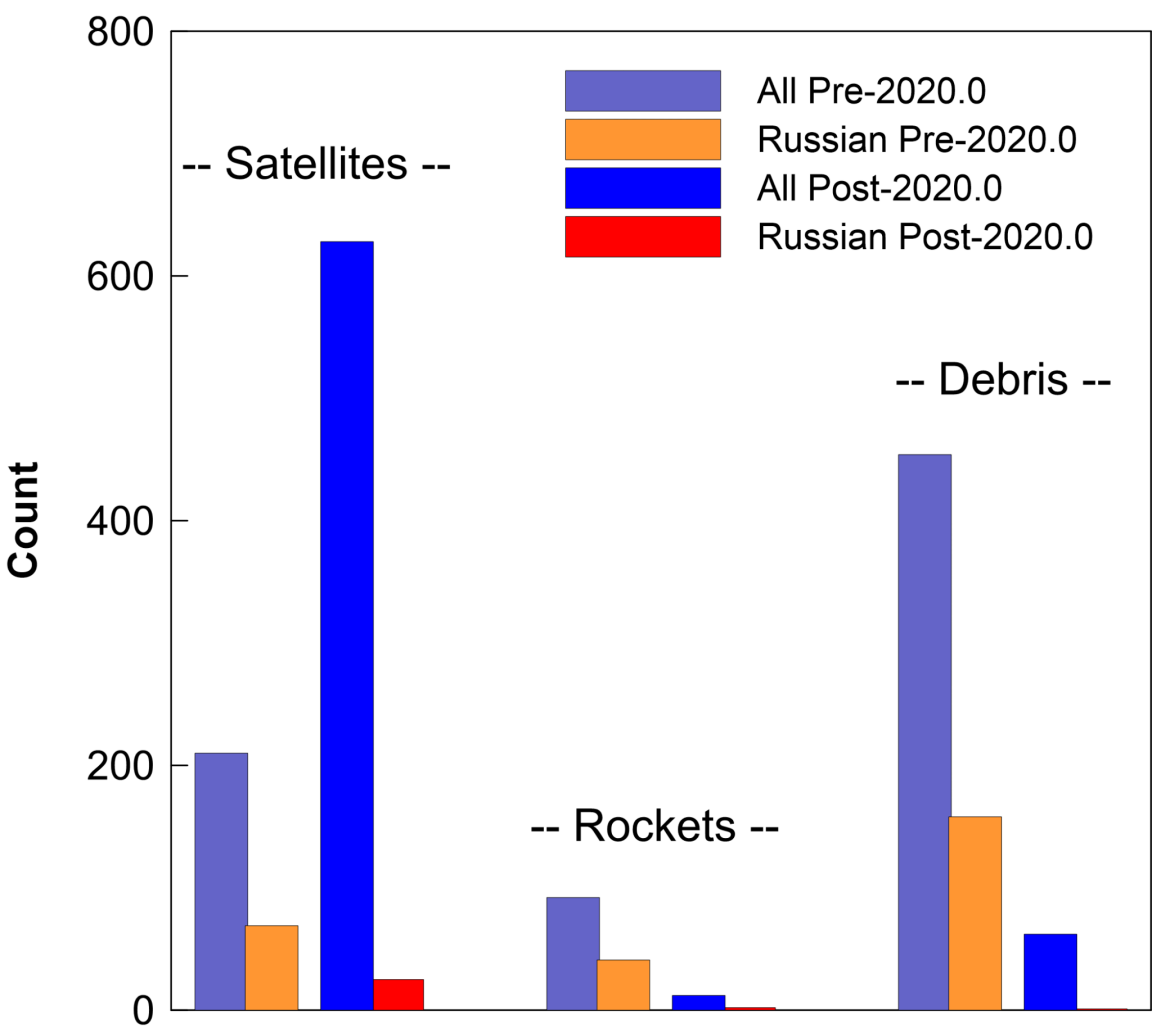

*Figure 1. Counts of the three groups of objects: satellites, rockets and debris. The four bars for each group represent all objects launched before 2020.0, Russian objects of that same early era, all objects launched after 2020.0 and Russian objects of that same late era.*

Notice that the number of each object type has changed significantly since 2020.0. Satellites have increased by a factor of 3 while rockets and debris have each decreased by about a factor of 7.

**5. Brightness distribution**

The MMT9 apparent magnitudes were sorted into bins and are plotted as bars in Figure 2 to show their distribution. The peak is at magnitude 8.5 and the decline at fainter magnitudes is a selection effect caused by instrument sensitivity. The decline at brighter magnitudes is real and it is due to fewer numbers of large objects.

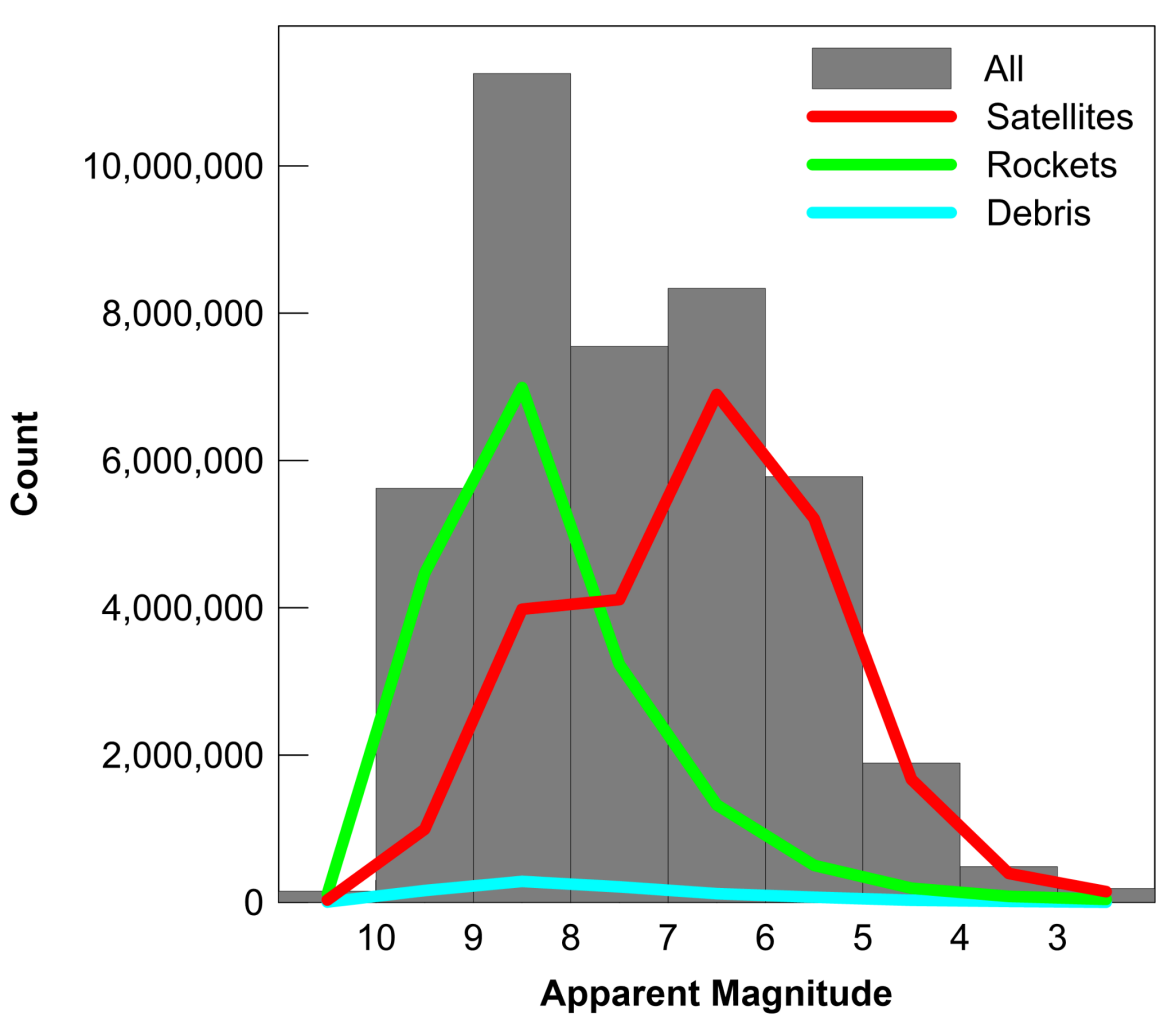

*Figure 2. Bars represent the distribution of apparent magnitudes for all objects. Lines are for the three types. Counts for satellites and rockets greatly outnumber those for debris objects.*

Observations of satellites are numerous and typically bright as shown by the red line. Those of rockets are also numerous but dimmer, and observations of debris objects are few in number and faint.

Examination of the light curves for many tracks of rapidly variable brightness objects indicates that MMT9 records them very reliably to magnitude 8, frequently to magnitude 9 and sometimes as faint as magnitudes 10 and 11. Since the trails of satellites fainter than magnitude 8 do not generally impact astronomical images, the data recorded by MMT9 is fully adequate for this research.

**6. Object densities on the sky**

Counts of observations with magnitudes brighter than 10.0 are listed in Table 2. In order to quantify their impact on astronomy, densities were computed for observations brighter than 6.0, 7.0 and 8.0 in units of objects per square degree of sky. This provides a snapshot view of their presence in the sky.

Table 2. Magnitude Counts

| Bin center | Count | Cumulative |
|---|---|---|
| -1.5 | 376 | 376 |
| -0.5 | 1,551 | 1,927 |
| 0.5 | 13,822 | 15,748 |
| 1.5 | 71,413 | 87,161 |
| 2.5 | 187,679 | 274,840 |
| 3.5 | 485,498 | 760,337 |
| 4.5 | 1,889,835 | 2,650,172 |
| 5.5 | 5,781,123 | 8,431,295 |
| 6.5 | 8,338,252 | 16,769,548 |
| 7.5 | 7,550,365 | 24,319,913 |
| 8.5 | 11,254,561 | 35,574,474 |
| 9.5 | 5,622,337 | 41,196,811 |

The field of view (FOV) for each MMT9 sensor is 99 square degrees, the total observing time for all 9 channels taken together during 2025 was 22.3 e6 seconds and observations were recorded at 10 Hz. Thus, the sky coverage for 2025 was 22.1 e9 square degrees. (Note: The different FOVs may overlap in which case a satellite in an area of overlap will be recorded in both channels.)

The cumulative number of observations brighter than magnitude 6.0, 7.0 and 8.0 are 8.4 e6, 16.8 e6 and 24.3 e6. So, their densities are 0.38 e-3, 0.76 e-3 and 1.10 e-3 objects per square degree of sky as listed in Table 3.

Table 3. Densities (objects / square degree)

| Mag Limit | Density |
|---|---|
| 6.0 | 0.00038 |
| 7.0 | 0.00076 |
| 8.0 | 0.00110 |

**7. Probability of appearing on an image**
In order to determine the probability that an artificial space object will appear on an astronomical image during the exposure, its angular rate of travel across the sky must be determined. That value depends on the object's altitude above the Earth and its height above the horizon. Lower altitude spacecraft travel faster due to the greater gravitational attraction, while those closer to the horizon appear slower because of their greater distances.

There are thousands of objects at different altitudes and they can appear anywhere in the sky. Thus, simplifying estimates are needed to make the computation of probabilities tractable. The altitudes of spacecraft in large constellations range from lows of 350 to 475 km for Starlink spacecraft to a high of 1,200 km for OneWeb. The majority are Starlink satellites, so we take the altitude to be 500 km which is toward the low end of the range. Astronomers generally prefer to observe targets when they are as near the zenith as possible. So, we take the height above the horizon to be 90°. The apparent speed of an object with that altitude and height is 0.88 degrees per second.

We evaluated probabilities for a range of exposure durations from 1 to 10 seconds; FOV sizes from 1 to 10 degrees square and magnitude limits of 7.0 and 8.0.

Probabilities were determined by examining a grid of sky cells of the same size as the FOV and centered on the FOV. The grid is made large enough to include all objects that could enter the FOV during the exposure.

The combined probability for multiple events equals 1.0 minus the product of the probabilities of non-occurrence for each individual event. In this case, that product is for the probabilities determined for each of the cells that an object from that cell did *not* enter the FOV.

For each cell in the grid 1,000 trials were run to determine whether an object within its boundary entered the FOV. A pseudo random number generator determined the object's starting location within the cell and its direction of motion.

For each cell, an array of 1,000 results from the trials was filled with 1.0 if the object did *not* enter the FOV. Otherwise, it was filled with 1.0 minus the product of density times the square of the FOV. That product is the likelihood that the cell actually had an object in it.

The sum of the trial array values was divided by 1,000 to give the mean probability that an object from that cell did *not* enter the FOV. The probabilities for all cells in the grid were then multiplied together and the product was subtracted from 1.0. The resulting probabilities that at least one object was in the FOV are listed in Table 4 as a function of magnitude, exposure duration and FOV size.

The least probabilities of 0.002 are for a 1 degree FOV exposed for 1 second. The greatest, 0.212, corresponds to an object brighter than magnitude 8 in a 10 degree square FOV during a 10 second exposure.

Probabilities increase more slowly than exposure duration for a given FOV size. The change is less than linear because objects from greater distances outside the FOV are less likely to enter that field due to the smaller range of directions that will cross it. Likewise, probabilities increase more slowly than the square areas of FOV for a given exposure duration. In this case, more distant objects are traveling too slowly to enter the FOV.

Table 4. Probabilities of Interference

**Magnitude 7**

| FOV (deg) | Exposure (s) 1 | 2 | 5 | 10 |
|---|---|---|---|---|
| 1 | 0.002 | 0.002 | 0.005 | 0.009 |
| 2 | 0.005 | 0.006 | 0.011 | 0.020 |
| 5 | 0.023 | 0.028 | 0.040 | 0.060 |
| 10 | 0.084 | 0.093 | 0.114 | 0.151 |

**Magnitude 8**

| FOV (deg) | Exposure (s) 1 | 2 | 5 | 10 |
|---|---|---|---|---|
| 1 | 0.002 | 0.003 | 0.007 | 0.013 |
| 2 | 0.007 | 0.009 | 0.017 | 0.028 |
| 5 | 0.033 | 0.039 | 0.055 | 0.087 |
| 10 | 0.120 | 0.129 | 0.165 | 0.212 |

Probabilities for intermediate values of exposure durations and FOV sizes not in the table can be approximated with Equation 3.

$$P = S * F ^ {1.6} * E ^ {0.4} \quad \text{Equation 3}$$

*F* is the FOV size, *E* is the exposure duration and S is a scale factor equal to 1.6 e-3 and 2.3 e-3 for magnitudes 7 and 8, respectively.

RMS differences between Equation 3 and the corresponding table of values vary widely from the small to the large extremes of FOV and exposures. For FOVs and exposures of 1 and 2 (that is, the upper left quadrant of the Table) the RMS is 0.00 for magnitude 7. Meanwhile, the RMS is 0.007 for FOVs and exposures of 5 and 10 (the lower right quadrant). Probabilities also depend on the distance of the Sun below the horizon as discussed in Section 10.

**8. Probability of being seen by eye**

Visual perception of an artificial space object is fundamentally different from the object being recorded on an image. To begin with, the eye's response to a scene is practically instantaneous. So, there is no equivalent to exposure duration. Furthermore, visual sensitivity at night is greatest in a peripheral annulus outside of the 'point of fixation' where the observer is looking. The inner and outer radii are 12 and 30 degrees, giving 2,375 square degrees for the area of the annulus. The probability of an object being superposed on the annulus is the product of its area multiplied by the density of objects brighter than magnitude 6.0. Table 3 lists that density as 0.00038. So the expected number of objects in the annulus is 0.90 and the corresponding Poisson probability of at least one object being seen is 59%.

As with imagery, the probability of visually sighting an object is a function of the Sun's distance below the horizon. Section 10 addresses that topic.

## 9. Satellite constellation era

The results described above pertain to all satellites, rockets and debris objects that are currently orbiting the Earth. Some of them were launched during the early years of the space age. Meanwhile, the population of objects in orbit has changed significantly during the era of satellite constellations which we take to start at 2020.0.

Figure 1 and Table 1 illustrate that change for each type of object. Specifically, the population of satellites from the pre-constellation era outnumbers rockets from the same era by a factor of about 2:1. Now satellites outnumber rockets by 50:1. This is due to launching many constellation satellites on a single rocket. Meanwhile, satellites are typically about two magnitudes more luminous than rockets as shown in Figure 2. (The figure also reveals that magnitude counts for debris objects are insignificant.)

The changing population numbers and the different luminosity characteristics of spacecraft and rockets leads to a brighter magnitude distribution for objects from the satellite constellations era. Figure 3 and Table 5 show that constellation era objects peak at magnitude 6.5 while the earlier ones peak at 8.5. This mimics the distribution of magnitudes for the satellites and rockets themselves illustrated in Figure 2.

## 10. Discussion

Three topics are addressed here. First is the variation of densities and probabilities during the nighttime hours. Second is forecasting the impact of space objects on astronomy in coming years. Third is research papers relating to this study.

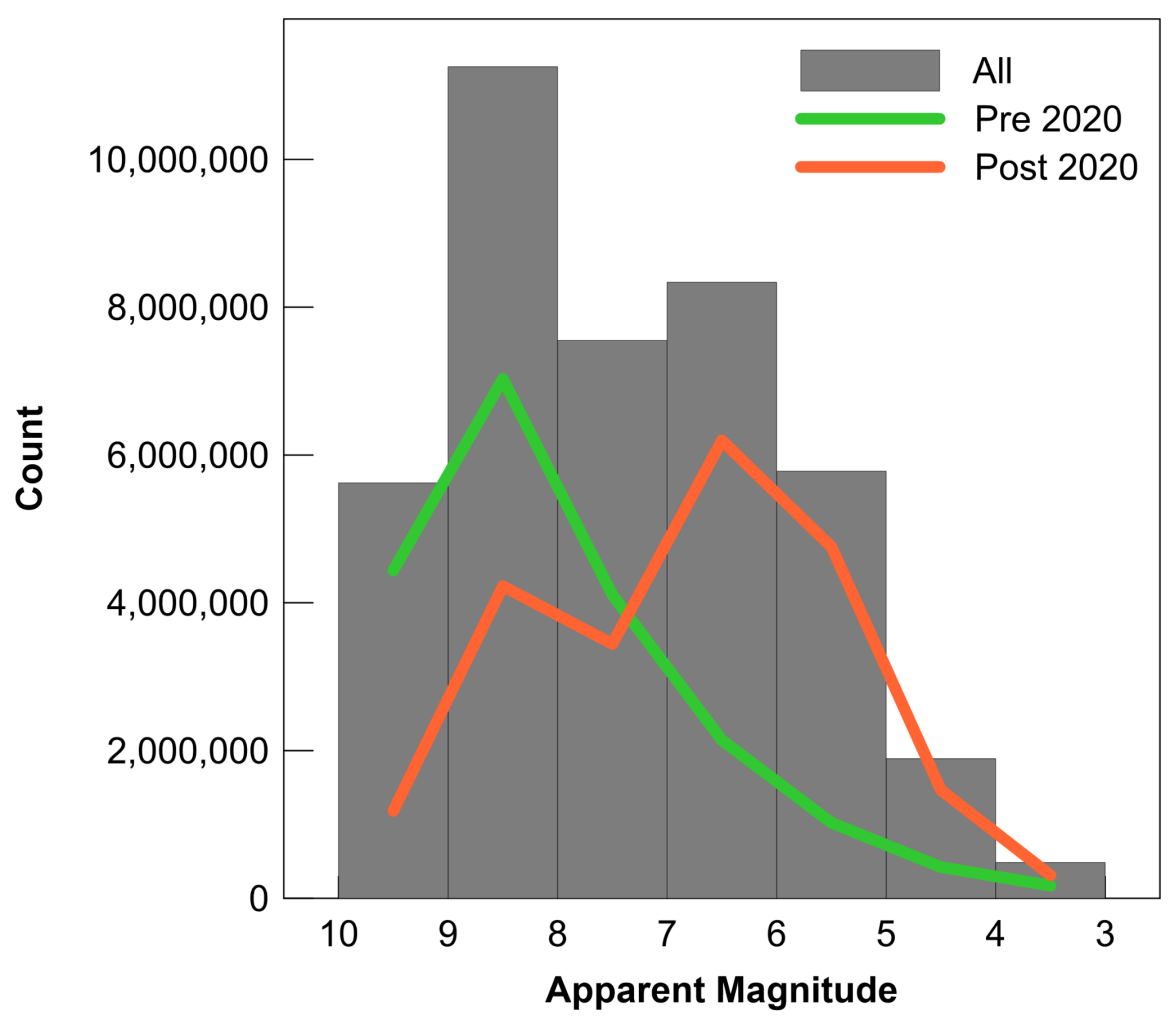

*Figure 3. Contributions to the magnitude distribution from objects launched before and after 2020.0.*

Table 5. Magnitude Counts by Time Period

| Bin Center | < 2020.0 | > 2020.0 |
|---|---|---|
| 3.5 | 171,952 | 313,546 |
| 4.5 | 423,578 | 1,466,257 |
| 5.5 | 1,023,432 | 4,757,691 |
| 6.5 | 2,141,356 | 6,196,896 |
| 7.5 | 4,107,080 | 3,443,285 |
| 8.5 | 7,029,417 | 4,225,144 |
| 9.5 | 4,438,561 | 1,183,776 |

## 10a. Variation during the nighttime hours

The Earth casts its shadow into the sky at night. When the Sun is at nadir that shadow is centered at zenith and it eclipses nearly all low Earth orbiting objects. Therefore, the fewest number of objects are visible at any location around the time of local midnight.

During twilight, on the other hand, the shadow is cast toward the horizon in the direction opposite the Sun’s azimuth. Then most objects in the sky are sunlit. Eclipsed sky regions for satellites at 500 km altitude are shown in Figure 4 for the Sun 10 and 30 degrees below the horizon. In the former case, just 6% of the sky is in the eclipse region, while 86% is in that region in the latter case.

The MMT9 observatory starts and finishes operation when the Sun is 10 degrees below the horizon. This solar elevation limit is typical for an astronomical observatory. So, the densities and probabilities derived from MMT9 observations in this paper are representative of what can be expected at other observatories. Those values will be higher during twilight and lower around midnight.

**10b. Forecast for coming years**

Satellites from the constellation era (2020.0 and onward) dominate the distribution of bright magnitudes as reported in Section 9. In 2025 there were about 10,000 such spacecraft in low Earth orbit, but satellite operators' projections for coming years put that number at more than one million.

If those plans for a 100-fold increase are realized, nearly all astronomical images of a few seconds duration and a few degrees in size will be compromised by the presence of satellite trails during much of each night. For example, the current probability of 4.0% for a 5 second exposure, 5 degree FOV and an object brighter than magnitude 7.0 (Table 4) increases to nearly 100%.

Likewise the density of objects brighter than magnitude 6.0 would increase by a factor of 100, and about 90 objects will be visible in the eye's peripheral vision annulus. By comparison, there are 5,326 stars brighter than magnitude 6.0 according to UCAC4. The corresponding density is 0.129 per square degree and the number in the annulus is 306. Thus, satellites will be about 29% as numerous as stars.

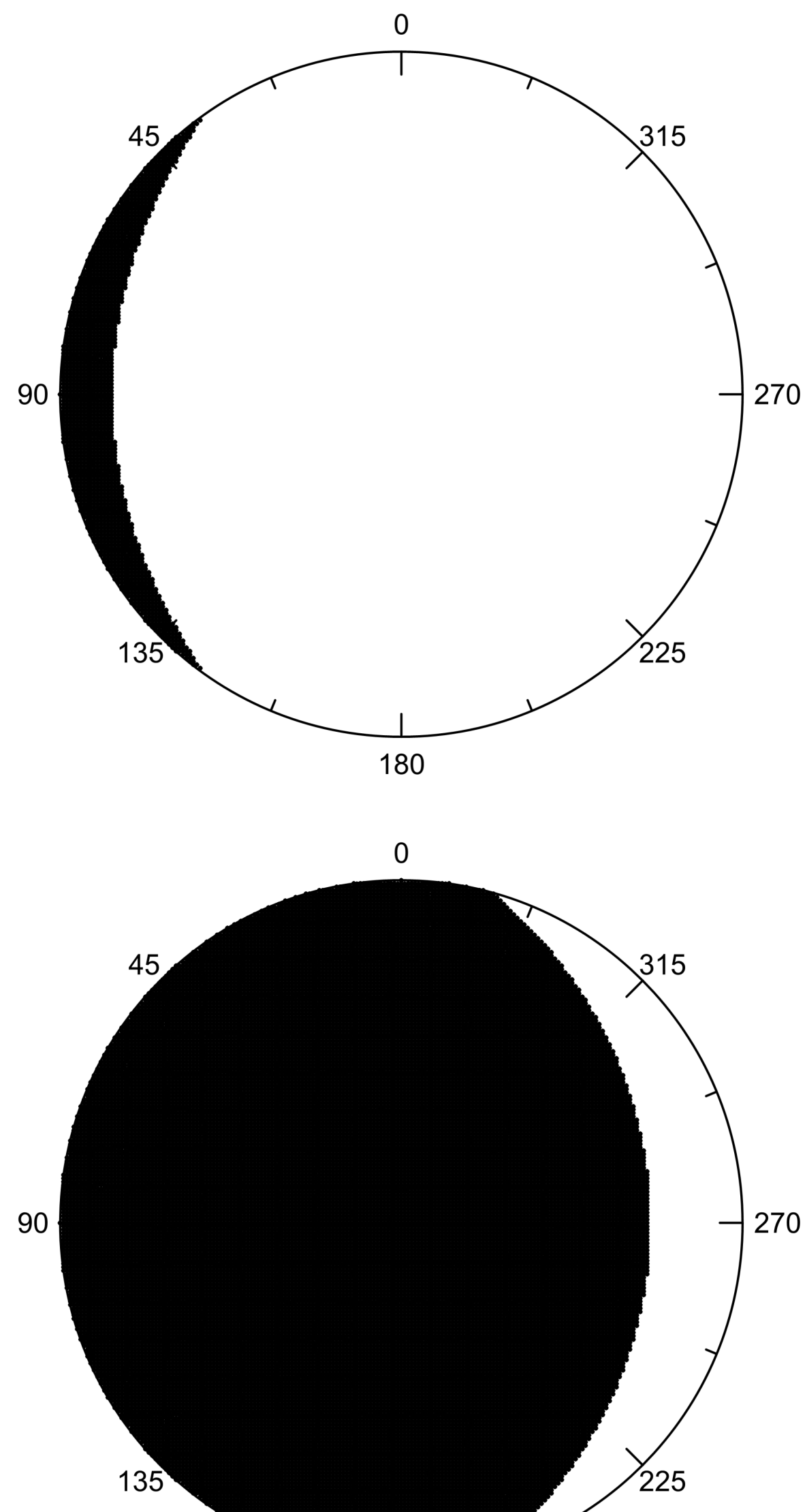

*Figure 4. Sky regions (black) where satellites at 500 km are eclipsed. The top panel is for the Sun 10 degrees below the horizon and bottom panel is for 30 degrees. The solar azimuth is 270 degrees. The sky is projected on a plane.*

Motion is a trigger for visual perception and it results in *attention capture*. Therefore, visual observers will perceive satellites more strongly than stars.

A caveat to these predictions is that many of the new spacecraft may be in orbits that are Sun-synchronous. Furthermore, linear scaling of density with satellite numbers assumes that the brightness distribution and altitudes of a future million-satellite population resembles the current scenario. Those factors would change the probabilities of their being imaged and seen.

**10c. Related research**

Hainaut (2026) performed a comprehensive study pertaining to the impact of large and bright satellite constellations due to satellite streaks on images, diffuse background, and scattered sky brightness. He simulated a mixture of spacecraft at altitudes from 328 to 1200 km and absolute magnitude 7 at 550 km. Hainaut determined that, "mega-constellations with $10^6$ satellites render trails pervasive, thereby affecting the majority of long exposures." The findings in this paper are consistent with his result.

The authors of this paper recently examined the impact of debris objects alone on astronomy (Mallama et al 2026b). The present study confirmed that the effect of debris is a relatively minor concern.

**11. Summary and conclusions**

This study examines 32 million observations recorded during 2025 by the MMT9 observatory. We characterize the brightness of satellites, rockets and debris objects. The densities of observations (per square degree of sky) bright enough to impact astronomy are reported. Those densities are used to compute probabilities that the objects will impact astronomical images and that objects will be visible to the unaided eye. The effect of the Sun's distance below the horizon is addressed.

Trends that have emerged during the era of satellite constellations are discussed. If plans to launch a million satellites are realized, nearly all astronomical images of a few seconds duration and a few degrees in size may be compromised during several hours each night. Finally, people viewing the night sky may notice satellites more than stars.

**Data availability**

The observations studied in this article are available from the Mini-MegaTORTORA (MMT9) online database. Software used for the analysis is available from the corresponding author.

**Acknowledgements**

M. Dickinson of the IAU-CPS conducted a review of this study. Comments and suggestions from two anonymous reviewers helped to improve the paper. D.R. Skillman independently reviewed the algorithm for computing probability.

The authors acknowledge the support of the International Astronomical Union (IAU) Centre for the Protection of the Dark and Quiet Sky (CPS, https://cps.iau.org). Any opinions, findings, and conclusions or recommendations expressed in this material are those of the author and do not necessarily reflect the views of the IAU, NSF NOIRLab, SKAO, ESO, or any host or member institution of the IAU CPS.

**References**

Barentine, J.C., Venkatesan, A., Heim, J., Lowenthal, J., Kocifa, M. and Bará, S. 2023. Aggregate effects of proliferating low-Earth-orbit objects and implications for astronomical data lost in the noise. Nature Astronomy, 7, 252-258. https://www.nature.com/articles/s41550-023-01904-2.

Beskin, G.M., Karpov, S.V., Biryukov, A.V., Bondar, S.F., Ivanov, E.A., Katkova, E.V., Orekhova, N.V., Perkov, A.V. and Sasyuk, V.V. 2017. Wide-field optical monitoring with Mini-MegaTORTORA (MMT-9) multichannel high temporal resolution telescope. Astrophysical Bulletin. 72, 81-92. doi.org/10.1134/S1990341317030105.

Hainaut, O.R. 2026, Large or bright satellite constellations: Effects on observations, including on the background sky brightness. https://arxiv.org/abs/2604.09427.

IAU Centre for the Protection of the Dark, Quiet Sky from Satellite Constellation Interference and 40 co-authors. 2024. Call to protect the dark and quiet sky from harmful interference by satellite constellations. https://arxiv.org/abs/2412.08244.

Karpov, S., Katkova, E., Beskin, G., Biryukov, A., Bondar, S., Davydov, E., Perkov, A. and Sasyuk, V. 2016. Massive photometry of low-altitude artificial satellites on Mini-MegaTORTORA. Revista Mexicana de Astronomía y Astrofísica (Serie de Conferencias) Vol. 48, pp. 112-113.

Longa-Peña, P. and 47 co-authors. 2026. Brightness evolution of LEO Starlink mega-constellation satellites from 2021 to 2023: a multiyear ground-based photometric study. MNRAS, 548-552L, doi.org/10.1093/mnras/stag552.

Mallama, A. 2021. Starlink Satellite brightness – characterized from 100,000 visible light magnitudes. https://arxiv.org/abs/2111.09735.

Mallama, A. and Young, M. 2021. The satellite saga continues. Sky and Telescope, **141**, June, p. 16.

Mallama, A. and Cole, R.E. 2025. Satellite constellations exceed the limits of acceptable brightness established by the IAU, MNRAS, 544, 1, L15-17, doi.org/10.1093/mnrasl/slaf09.

Mallama, A., Cole, R.E., Zhi, H., Young, B., Respler, J., Zamora, O. and Dadighat, M. 2026a. Brightness characterization and modeling for Amazon Leo satellites. https://arxiv.org/abs/2601.07708.

Mallama, A., Karpov, S. and Cole, R.E. 2026b. The impact of space debris on optical astronomy. https://arxiv.org/abs/2607.05480.

Zhi, H., Jiang, X., and Wang, J. Multicolour photometry of LEO mega-constellations Starlink and OneWeb, MNRAS, 530, 5006–5015, https://doi.org/10.1093/mnras/stae693.